\documentclass[12pt]{article}
\usepackage{amsmath}
\usepackage[dvips]{graphicx}
\input amssym

\DeclareMathOperator{\diag}{diag}
\begin{document}

\def\prg#1{\medskip\noindent{\bf #1}}  \def\ra{\rightarrow}
\def\lra{\leftrightarrow}              \def\Ra{\Rightarrow}
\def\nin{\noindent}                    \def\pd{\partial}
\def\dis{\displaystyle}                \def\inn{\hook}
\def\grl{{GR$_\Lambda$}}               \def\Lra{{\Leftrightarrow}}
\def\cs{{\scriptstyle\rm CS}}          \def\ads3{{\rm AdS$_3$}}
\def\Leff{\hbox{$\mit\L_{\hspace{.6pt}\rm eff}\,$}}
\def\bull{\raise.25ex\hbox{\vrule height.8ex width.8ex}}
\def\ric{{(Ric)}}                      \def\tric{{(\widetilde{Ric})}}
\def\tmgl{\hbox{TMG$_\Lambda$}}
\def\Lie{{\cal L}\hspace{-.7em}\raise.25ex\hbox{--}\hspace{.2em}}
\def\sS{\hspace{2pt}S\hspace{-0.83em}\diagup}   \def\hd{{^\star}}
\def\dis{\displaystyle}                 \def\ul#1{\underline{#1}}

\def\hook{\hbox{\vrule height0pt width4pt depth0.3pt
\vrule height7pt width0.3pt depth0.3pt
\vrule height0pt width2pt depth0pt}\hspace{0.8pt}}
\def\semidirect{\;{\rlap{$\supset$}\times}\;}
\def\first{\rm (1ST)}       \def\second{\hspace{-1cm}\rm (2ND)}
\def\bm#1{\hbox{{\boldmath $#1$}}}
\def\nb#1{\marginpar{{\large\bf #1}}}

\def\G{\Gamma}        \def\S{\Sigma}        \def\L{{\mit\Lambda}}
\def\D{\Delta}        \def\Th{\Theta}
\def\a{\alpha}        \def\b{\beta}         \def\g{\gamma}
\def\d{\delta}        \def\m{\mu}           \def\n{\nu}
\def\th{\theta}       \def\k{\kappa}        \def\l{\lambda}
\def\vphi{\varphi}    \def\ve{\varepsilon}  \def\p{\pi}
\def\r{\rho}          \def\Om{\Omega}       \def\om{\omega}
\def\s{\sigma}        \def\t{\tau}          \def\eps{\epsilon}
\def\nab{\nabla}      \def\btz{{\rm BTZ}}   \def\heps{\hat\eps}
\def\bu{{\bar u}}     \def\bv{{\bar v}}     \def\bs{{\bar s}}
\def\bx{{\bar x}}     \def\by{{\bar y}}     \def\bom{{\bar\om}}
\def\tphi{{\tilde\vphi}}  \def\tt{{\tilde t}}

\def\tG{{\tilde G}}   \def\cF{{\cal F}}      \def\bH{{\bar H}}
\def\cL{{\cal L}}     \def\cM{{\cal M }}     \def\cE{{\cal E}}
\def\cH{{\cal H}}     \def\hcH{\hat{\cH}}
\def\cK{{\cal K}}     \def\hcK{\hat{\cK}}    \def\cT{{\cal T}}
\def\cO{{\cal O}}     \def\hcO{\hat{\cal O}} \def\cV{{\cal V}}
\def\tom{{\tilde\omega}}                     \def\cE{{\cal E}}
\def\cR{{\cal R}}    \def\hR{{\hat R}{}}     \def\hL{{\hat\L}}
\def\tb{{\tilde b}}  \def\tA{{\tilde A}}     \def\tv{{\tilde v}}
\def\tT{{\tilde T}}  \def\tR{{\tilde R}}     \def\tcL{{\tilde\cL}}
\def\hy{{\hat y}\hspace{1pt}}  \def\tcO{{\tilde\cO}}

\def\nn{\nonumber}                    \def\vsm{\vspace{-9pt}}
\def\be{\begin{equation}}             \def\ee{\end{equation}}
\def\ba#1{\begin{array}{#1}}          \def\ea{\end{array}}
\def\bea{\begin{eqnarray} }           \def\eea{\end{eqnarray} }
\def\beann{\begin{eqnarray*} }        \def\eeann{\end{eqnarray*} }
\def\beal{\begin{eqalign}}            \def\eeal{\end{eqalign}}
\def\lab#1{\label{eq:#1}}             \def\eq#1{(\ref{eq:#1})}
\def\bsubeq{\begin{subequations}}     \def\esubeq{\end{subequations}}
\def\bitem{\begin{itemize}}           \def\eitem{\end{itemize}}
\renewcommand{\theequation}{\thesection.\arabic{equation}}
\title{Vaidya-like exact solutions with torsion}

\author{M. Blagojevi\'c and B. Cvetkovi\'c\footnote{
        Email addresses: {mb@ipb.ac.rs, cbranislav@ipb.ac.rs}} \\
Institute of Physics, University of Belgrade,\\
                      P. O. Box 57, 11001 Belgrade, Serbia}
\date{\today}
\maketitle

\begin{abstract}
Starting from the Oliva--Tempo--Troncoso black hole, a solution of the
Bergshoeff--Hohm--Townsend massive gravity, a class of the Vaidya-like
exact vacuum solutions with torsion is constructed in the
three-dimensional Poincar\'e gauge theory. A particular subclass of
these solutions is shown to possess the asymptotic conformal sym\-metry.
The related canonical energy contains a contribution stemming from
torsion.
\end{abstract}

\section{Introduction}
\setcounter{equation}{0}

Poincar\'e gauge theory (PGT) is a modern field-theoretic approach to
gravity, proposed in the early 1960s by Kibble and Sciama \cite{x1}.
Compared to Einstein's general relativity (GR), PGT is based on using both
the torsion and the curvature to describe the underlying Riemann--Cartan
(RC) geometry of spacetime \cite{x2,x3,x4}. Investigations of PGT in
three-dimensional (3D) spacetime are expected to improve our understanding
of both the geometric and dynamical role of torsion in a realistic,
four-dimensional gravitational theory. Systematic studies of 3D PGT
started  with the Mielke--Baekler model \cite{x5}, introduced in the 1990s
as a PGT extension of GR. However, this model is, just like GR, a
topological theory without propagating degrees of freedom. In PGT, such an
unrealistic dynamical feature can be quite naturally improved by going
over to Lagrangians that are quadratic in the field strengths
\cite{x6,x7}, as in the standard gauge theories.

Relying on our experience with GR, we know that exact solutions of a
gravitational theory are essential for its physical interpretation. In the
context of 3D PGT, exact solutions were first studied in the
Mielke--Baekler model; for a review, see Chapter 17 in Ref. \cite{x4}.
Recently, our research interest moved toward exact solutions in a more
dynamical framework of the quadratic PGT. The first step in this direction
was made by constructing the Ba\~nados--Teitelboim-Zanelli (BTZ) black
hole with torsion \cite{x7}. Then, we showed that gravitational waves can
be naturally incorporated into the PGT dynamical framework \cite{x8,x9}.
The purpose of the present work is to examine a PGT generalization of the
Oliva--Tempo--Troncoso (OTT) black hole \cite{x10}, see also \cite{x11},
as well as its Vaidya-like extension \cite{x12}.

The OTT black hole is an exact solution of the Bergshoeff--Hohm--Townsend
(BHT) massive gravity \cite{x13}, a Riemannian model defined by adding a
specific combination of curvature-squared terms to the Hilbert--Einstein
action. Generically, the BHT gravity with a cosmological constant admits
two distinct maximally symmetric vacua. However, when the coupling
constants satisfy a specific critical condition, these two vacua coincide.
It is exactly in this case that the OTT black hole is a vacuum solution of
the BHT gravity.\footnote{For the canonical aspects of the full nonlinear
theory in the critical regime, see Refs. \cite{x16,x17}.} Going a step
further, Maeda \cite{x12} formulated a Vaidya-like extension of the OTT
black hole, assuming the presence of a null dust fluid as a \emph{matter
field}. In this paper, we construct a Vaidya-OTT spacetime with torsion as
an exact \emph{vacuum solution} of PGT.

The paper is organized as follows. In section 2, we describe the static
OTT black hole as a Riemannian solution of PGT in vacuum. In particular,
the canonical expression for the gravitational energy is shown to be
directly compatible with the first law of black hole thermodynamics. In
section 3, we introduce a Vaidya extension of the OTT metric in the manner
of Maeda \cite{x12}; the resulting Riemannian geometry is not compatible
with the PGT dynamics in vacuum. Then, in section 4, we construct a
Vaidya--OTT geometry with torsion as an exact vacuum solution of PGT. In
section 5, we apply canonical methods to show that a specific subclass of
these solutions is characterized by the asymptotic conformal symmetry. The
canonical Vaidya--OTT energy is found; apart from the OTT term, it
contains a contribution stemming from torsion. The associated surface term
of the canonical generator for time translations is a generalization of
the more standard expression \cite{x14}, used in Ref. \cite{x15} to
calculate energies for a number of exact solutions in 3D gravity. Finally,
section 6 is devoted to concluding remarks, while appendices contain some
technical details.

Working in PGT, we use the following conventions: the Latin indices $(i,
j, k, ...)$ refer to the local Lorentz frame, the Greek indices
$(\m,\n,\r, ...)$ refer to the coordinate frame, $b^i$ is the triad field
(coframe 1-form), $\om^{ij}=-\om^{ji}$ is a connection 1-form, the
respective field strengths are the torsion $T^i=db^i+\om^i{_m}\wedge b^m$
and the curvature $R^{ij}=d\om^{ij}+\om^i{_k}\wedge \om^{kj}$ (2-forms);
the Lie dual of an antisymmetric form $X^{ij}$ is
$X_i:=-\ve_{ijk}X^{jk}/2$, the Hodge dual of a form $\a$ is $\hd\a$, and
the exterior product of forms is implicit.

\section{OTT black hole in PGT}
\setcounter{equation}{0}

We begin our considerations by showing that the static OTT black hole, a
vacuum solution of the BHT gravity with a unique AdS ground state
\cite{x10}, is also a \emph{Riemannian solution} of PGT, in spite of the
fact that PGT represents quite a different dynamical framework \cite{x7}.

\subsection{Geometric aspects}

The metric of the static OTT spacetime is given by
\be
ds^2=N^2dt^2-\frac{dr^2}{N^2}-r^2 d\vphi^2\, , \qquad
N^2:=-\m+Br+\frac{r^2}{\ell^2}\, ,                              \lab{2.1}
\ee
where $\m$ and $B$ are integration constants. The Killing horizons are
determined by the condition $N^2=0$:
$$
r_\pm=\frac{\ell^2}{2}\left(-B\pm\sqrt{B^2+4\m/\ell^2}\right)\,.
$$
When at least $r_+$ is real and positive, and $\ell^2>0$, the OTT metric
defines a static and spherically symmetric AdS black hole; for $B=0$, it
reduces to the BTZ black hole.

Given the metric \eq{2.1}, one can choose the associated triad field in
the form
\bea
b^0=Ndt\, ,\qquad b^1=\frac{dr}{N}\, ,\qquad b^2=rd\vphi\, ,    \lab{2.2}
\eea
so that $ds^2=\eta_{ij}b^i\otimes b^j$, where $\eta=\diag(+1,-1,-1)$.
Then, treating the OTT black hole as a Riemannian object, we use the
Christoffel connection,
\be
\om^{01}=-N'b^0\, , \qquad \om^{02}=0\, ,\qquad
\om^{12}=\frac{N}{r}b^2\, ,                                     \lab{2.3}
\ee
where $N':=\pd N/\pd r$, to calculate the curvature 2-form:
\bsubeq\lab{2.4}
\bea
&&R^{01}=(N'N)'b^0b^1=\frac{1}{\ell^2}b^0b^1\, ,                \nn\\
&&R^{02}=\frac{1}{r}N'Nb^0b^2
        =\left(\frac{B}{2r}+\frac{1}{\ell^2}\right)b^0b^2\, ,   \nn\\
&&R^{12}=\frac{1}{r}N'Nb^1b^2
        =\left(\frac{B}{2r}+\frac{1}{\ell^2}\right)b^1b^2\, .
\eea
For $B\ne 0$, the scalar curvature has a singularity at $r=0$:
\be
R=\frac{6}{\ell^2}+\frac{2B}{r}\, .
\ee
\esubeq
Nonvanishing irreducible components of the curvature are:
$$
{}^{(6)}R^{ij}=\frac{1}{6}Rb^ib^j\, ,\qquad
{}^{(4)}R^{ij}=R^{ij}-{}^{(6)}R^{ij}\, .
$$
In this geometry, the Cotton 2-form $C^i$, defined by
\be
C^i:=\nab L^i\,,\qquad L^i=\ric^i-\frac{1}{4}Rb^i\, ,           \lab{2.5}
\ee
is vanishing, so that the OTT spacetime is conformally flat. This is not a
surprise since the OTT metric is also a solution of the conformal gravity
\cite{x11}.

Now, we shall show that the OTT black hole is a Riemannian solution of
PGT in vacuum.

\subsection{Riemannian sector of PGT}

Lagrangian dynamics of PGT if expressed in terms of its basic field
variables, the triad field $b^i$ and the RC connection $\om^{ij}$, and the
related field strengths, the torsion $T^i$ and the curvature $R^{ij}$. The
general parity-preserving gravitational Lagrangian of PGT is quadratic in
the field strengths, see Appendix A. In the Riemannian sector of PGT,
torsion vanishes and $L_G$ is expressed only in terms of the curvature.
For ${}^{(5)}R_{ij}=0$, we have
\be
L_G=-\hd(a_0R+2\L_0)
    +\frac{1}{2}R^{ij}\hd\left(b_4{}^{(4)}R_{ij}
                              +b_6{}^{(6)}R_{ij}\right)\, ,     \lab{2.6}
\ee
and the general vacuum PGT field equations \eq{A.2} reduce to a simpler
form:
\bea
(1ST)&& E_i=0\, ,                                               \nn\\
(2ND)&& \nab H_{ij}=0\, ,                                       \lab{2.7}
\eea
where $E_i$ and $H_{ij}$ are given in \eq{A.5}. The field equations
produce the following result:
\bea
(2ND)&&\Ra\quad b_4+2b_6=0\, ,                                  \nn\\
(1ST)&&\Ra\quad b_4-2a_0\ell^2=0\,,\qquad a_0+2\ell^2\L_0=0\, . \lab{2.8}
\eea
Thus, the OTT black hole is an exact vacuum solution in the Riemannian
sector of PGT, provided the four Lagrangian parameters
$(a_0,b_4,b_6,\L_0)$ satisfy the above three conditions.

\subsection{Gravitational energy and entropy}\label{sub23}

Asymptotically, for large $r$, the OTT geometry takes the AdS form. Based
on the canonical approach described in Appendix B and section \ref{sec5},
one finds that the only nontrivial conserved charge of this geometry is the
gravitational energy,
\be
E=\frac{1}{4G}\left(\mu+\frac{1}{4}B^2\ell^2\right)\, ,         \lab{2.9}
\ee
whereas the angular momentum $M$ vanishes. The result is obtained from the
canonical generator of time translations, the surface term of which
contains a new contribution with respect to the more standard situation,
see Refs. \cite{x14,x15} and subsection \ref{sub52}.

Remarkably, the canonical expression for $E$ is directly compatible with
the first law of black hole thermodynamics. Indeed, using the OTT central
charges (subsection \ref{sub53})
\be
c^\pm=24\pi\cdot 2a_0\ell=\frac{3\ell}{G}\, ,
\ee
the Cardy formula produces the following expression for the entropy:
\be
S=4\pi\ell\sqrt{E/4G}\, .
\ee
Then, by introducing the Hawking temperature,
\be
T=\frac{1}{4\pi}\left.\pd_r N^2\right|_{r=r_+}
 =\frac{1}{2\pi\ell}\sqrt{4GE}\, ,
\ee
one can directly verify the first law of the black hole thermodynamics:
\be
\d E=T\d S\,.                                                   \lab{2.13}
\ee
Since the entropy vanishes for $E=0$, the state with $E=0$ can be
naturally regarded as the ground state of the OTT family of black holes
\cite{x18}.

The canonical energy \eq{2.9} coincides with the shifted OTT energy $\D
M=M-M_0$, introduced by Giribet et al. \cite{x18}, where $M=\mu/4G$ is
interpreted as the conserved charge and $M_0=-B^2/16 G$. The quantity $\D
M$ is defined to respect Cardy's formula for the entropy, and it has the
role of thermodynamic energy in the first law. In the canonical approach,
the conserved charge $E$ is the same object as the thermodynamic energy.

\section{Vaidya extension of the OTT metric}
\setcounter{equation}{0}

To obtain a Vaidya extension of the OTT metric, we first make a coordinate
transformation from the Schwarzschild-like time coordinate $t$ to a new
coordinate $u$, such that
\be
dt=du+dr/N^2\, .
\ee
The physical meaning of $u$ is obtained by noting that $u=$ const.
corresponds to a radially outgoing null ray, $dr/dt=N^2$, see Ref.
\cite{x19}. Then, following Maeda \cite{x12}, we introduce a Vaidya
extension of the OTT black hole by making $B$ a function of $u$, $B=B(u)$,
but leaving $\m$ as a constant. The Vaidya--OTT metric defines a time
dependent spherically symmetric geometry:
\be
ds^2=N^2du^2+2dudr-r^2d\vphi^2\, .
\ee

In the new coordinates $x^\m=(u,r,\vphi)$, it is convenient to choose the
triad field as
\be
b^+:=du\, , \qquad b^-:=Hdu+dr\, ,\qquad b^2:=rd\vphi\, ,       \lab{3.3}
\ee
where $H=N^2/2$, so that the line element becomes $ds^2=\eta_{ij}b^ib^j$,
with
$$
\eta_{ij}=\left( \ba{ccc}
                0  & 1 &  0   \\
                1  & 0 &  0   \\
                0  & 0 & -1
               \ea\right)\, .
$$
The dual frame $h_i$, defined by $h_i\inn b^j=\d^j_i$, is given by
$$
h_+=\pd_u-H\pd_r\,,\qquad h_-=\pd_r\, ,\qquad h_2=\frac{1}{r}\pd_\vphi\,.
$$

For vanishing torsion, one can use the Riemannian connection
\be
\om^{+-}=-H'b^+\, ,\qquad \om^{+2}=-\frac{1}{r}b^2\, ,\qquad
\om^{-2}=\frac{1}{r}Hb^2\, ,                                    \lab{3.4}
\ee
to calculate the related curvature 2-form $R^{ij}$. Then, following the
procedure described in the previous section, one finds that the PGT field
equations \eq{2.7} imply:
\bea
(2ND)&&\Ra\quad b_4+2b_6=0\, ,                                  \nn\\
(1ST)&&\Ra\quad b_4-2a_0\ell^2=0\, ,\qquad a_0+2\ell^2\L=0\, ,
       \qquad \ul{\dot B=0}\, ,                                 \lab{3.5}
\eea
where $\dot B:=\pd_u B$. Thus, the Vaidya--OTT metric with $\dot B\ne 0$
is \emph{not a Riemannian solution of PGT in vacuum}.

In order to overcome a similar barrier in the BHT gravity, Maeda
\cite{x12} introduced the Vaidya-OTT solution in the presence of
\emph{matter}, represented by a null dust fluid. The energy density of
this fluid is expressed directly in terms of the metric function $B(u)$,
which remains dynamically undetermined. Based on our experience with exact
wave solutions in PGT \cite{x8,x9}, we expect that the presence of torsion
could lead to a consistent description of the Vaidya-OTT dynamics \emph{in
vacuum}. Further exposition confirms this expectation.

\section{Vaidya--OTT solution with torsion}
\setcounter{equation}{0}

\subsection{Geometry of the ansatz}\label{sub41}

Following the logic of our approach to exact wave solutions in PGT
\cite{x8,x9}, we propose to look for a Vaidya--OTT solution with torsion
using the following two assumptions:
\bitem
\item[(i)] The new triad field retains the form \eq{3.3};\vsm
\item[(ii)] The RC connection is obtained from the Riemannian expression
    \eq{3.4} by the rule $H\to H+K$, where $K=K(u)$:
\be
\om^{+-}=-H'b^+\, , \qquad \om^{+2}=-\frac{1}{r}b^2\, ,\qquad
\om^{-2}=\frac{1}{r}(H+K)b^2\, .                                \lab{4.1}
\ee
\eitem
The new function $K$ is expected to \emph{compensate} the presence of the
problematic $\dot B$ term in the Riemannian field equations \eq{3.5}.
Geometrically, $K$ defines the torsion of spacetime. Indeed, using
$T^i:=\nab b^i$ one obtains:
\be
T^+,T^-=0\, ,\qquad T^2=\frac{1}{r}Kb^+b^2\, .                  \lab{4.2}
\ee
The nonvanishing irreducible components of the torsion are ${}^{(1)}T^i$
and ${}^{(2)}T^i$.

To complete the geometric description of our ansatz, we use the connection
\eq{4.1} to calculate the RC curvature 2-form:
\bea
&&R^{+-}=H''b^+b^-=\frac{1}{\ell^2}b^+b^-\, ,                   \nn\\
&&R^{+2}=\frac{1}{r}H'b^+b^2
        =\left(\frac{1}{\ell^2}+\frac{B}{2r}\right)b^+b^2\, ,   \nn\\
&&R^{-2}=\frac{1}{r}H'b^-b^2
         +\frac{1}{r}\left(\dot H+\dot K+H'K\right)b^+b^2\, .   \lab{4.3}
\eea
For $B\ne 0$, the scalar curvature is singular at $r=0$:
$$
R=\frac{6}{\ell^2}+\frac{2B}{r}\, .
$$
The nonvanishing irreducible components of the curvature are
${}^{(6)}R^{ij}$ and ${}^{(4)}R^{ij}=R^{ij}-{}^{(6)}R^{ij}$.

With the adopted geometric structure of our ansatz, the general PGT
Lagrangian \eq{A.1} becomes effectively of the form
\bea
L_G&=&-\hd(a_0R+2\L_0)+T^i\hd(a_1{}^{(1)}T_i+a_2{}^{(2)}T_i)  \nn\\
   &&+\frac{1}{2}R^{ij}\,\hd(b_4{}^{(4)}R_{ij}+b_6{}^{(6)}R_{ij})\, .\lab{4.4}
\eea

\subsection{Solutions}

With a given geometry of our ansatz, we now wish to find the metric
function $H$ and the torsion function $K$ as solutions of the vacuum PGT
field equations \eq{A.2}. To ensure a smooth limit to the standard OTT
black hole for $B\to$ const., we impose the conditions \eq{2.8} on the
Lagrangian parameters. Then, the field equations \eq{A.2} take the form
\bea
(2ND)&& 2\dot K+BK=0\, , \qquad a_1,a_2=0\, ,                    \nn\\
(1ST)&& \dot B\ell^2+2K=0\, .                                    \lab{4.5}
\eea
The conditions $a_1,a_2=0$ effectively eliminate the $T^2$ terms from the
Lagrangian. Moreover, the second term in $R^{-2}$ vanishes on-shell. Such
a reduction of $R^{ij}$ to its OTT form (with $\dot H,K=0$) is a
manifestation of the compensating role of the torsion function $K$.

By combining the above two equations, one obtains
\be
2K-\frac{1}{4}B^2\ell^2=-K_0\ell^2\, ,\qquad
\dot B+\frac{1}{4}B^2=K_0\, ,
\ee
where $K_0$ is an integration constant, the first integral of the field
equations \eq{4.5}. Introducing a new constant $E$ by $K_0\ell^2=4GE-\m$,
the first equation takes the form
\be
4GE=\m+\frac{1}{4}B^2\ell^2-2K\, ,                              \lab{4.7}
\ee
where $E$ is recognized as a RC generalization of the gravitational energy
\eq{2.9}. The conservation law of $E$ is defined with respect to the
evolution along $u$, $dE/du=0$. However, $dt=du+dr/N^2$ implies
$t=u+\cO_1$, so that asymptotically, one expects $E$ to be conserved also
with respect to the Schwarzschild-like time $t$. In the next section, this
argument is confirmed by canonical methods.

Depending on the value of $K_0$, there exist three branches of solutions.

\prg{1.} $K_0=C_1^2$. Apart from the trivial case $B=2C_1$, $K=0$, one
finds:
\be
B=2C_1\tanh\frac{C_1}{2}(u+C_2)\, ,\qquad
K=-\frac{C_1^2\ell^2}{2\cosh^2\frac{C_1}{2}(u+C_2)}\, .         \lab{4.8}
\ee

\prg{2.} $K_0=-C_1^2$. By replacing $C_1\to iC_1$ in the solution \eq{4.8},
one obtains:
\be
B=-2C_1\tan\frac{C_1}{2}(u+C_2)\, ,\qquad
K=\frac{C_1^2\ell^2}{2\cos^2\frac{C_1}{2}(u+C_2)}\, .
\ee

\prg{3.} $K_0=0$.
\be
B=\frac{4}{u+C_2}\, ,\qquad K=\frac{2\ell^2}{(u+C_2)^2}\,.
\ee
The solutions in branches 2 and 3 are singular at finite values of $u$,
whereas the solutions in branch 1 are perfectly regular, and physically
most appealing.

\begin{figure}[hth]
\centering
\includegraphics[height=4cm]{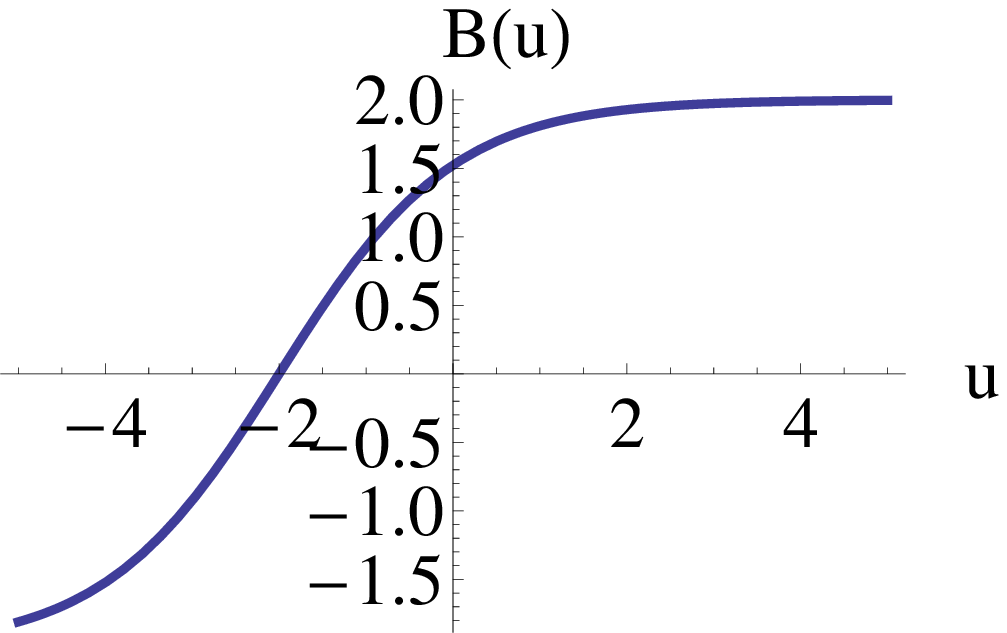}\qquad
\includegraphics[height=4cm]{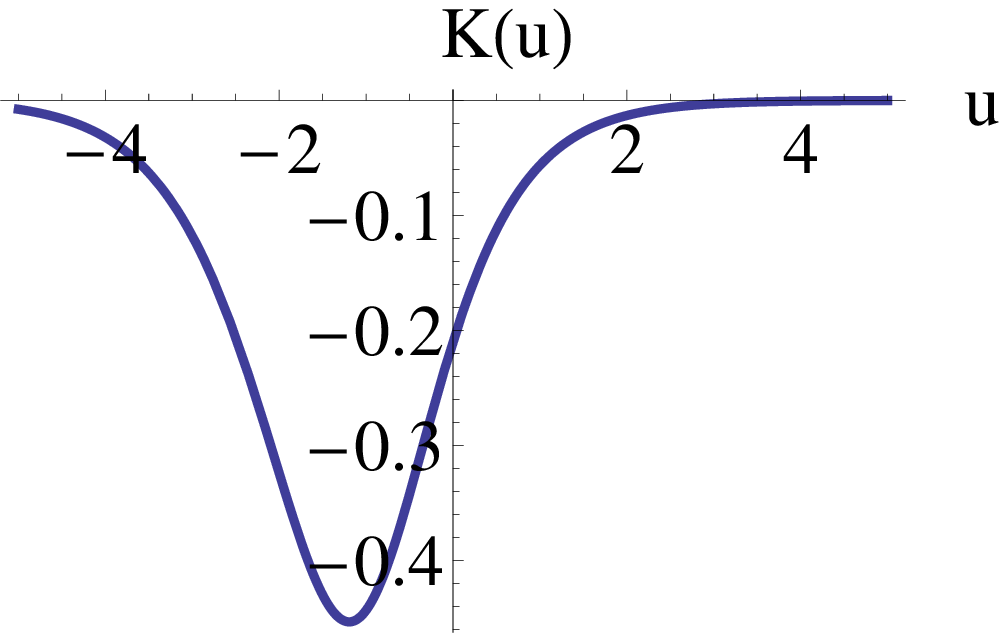}
\caption{Branch 1 solutions for $B(u)$ and $K(u)$, with
$C_1,\ell=1,C_2=2$.}
\end{figure}

In Figure 1, we illustrate a typical form of the solutions from branch 1.
Since $B(u)$ and $K(u)$, as well as their derivatives, are bounded
functions, the field strengths \eq{4.2} and \eq{4.3} approach
asymptotically to a Riemannian AdS spacetime. This motivates us to examine
the corresponding asymptotic structure in more details.

\section{Asymptotic symmetry}\label{sec5}
\setcounter{equation}{0}

In this section, we use the canonical approach to analyze the asymptotic
symmetry associated to the Vaidya--OTT solution with torsion in branch 1.

\subsection{AdS asymptotic conditions}

Transition from the OTT to the Vaidya--OTT triad is realized not only by
making $B$ a function of $u$, but also by going over to a new triad basis,
as can be seen by comparing Eqs. \eq{2.2} and \eq{3.3}. The new basis
allowed us to introduce the RC geometry by the simple rules formulated in
subsection \ref{sub41}. Then, requiring the \emph{invariance under the AdS
group} $SO(2,2)$, see \cite{x20}, one arrives at the following set of the
Vaidya--OTT asymptotic states:
\bsubeq\lab{5.1}
\be
b^i{_\m}=\bar b^i{_\m}+B^i{_\m},
\qquad
\bar b^i{_\m}=\left( \ba{lll}
            1    &   0   &  0             \\[1pt]
        \dis\frac{r^2}{2\ell^2}  & 1 & 0  \\[7pt]
            0    &  0    & r
               \ea\right),
\qquad
B^i{_\m}:=\left( \ba{lll}
         \cO_1    & \cO_3 & \cO_1  \\
         \cO_{-1} & \cO_1 & \cO_{-1}  \\
         \cO_0    & \cO_2 & \cO_0
               \ea\right),
\ee
and
\be
\om^i{_\m}=\bar\om^i{_\m}+\Om^i{_\m},
\qquad
\bom^i{_\m}=-\left( \ba{lll}
        0  &  0 &  1                      \\
        0  &  0 & \dis\frac{r^2}{2\ell^2} \\
   \dis\frac{r}{\ell^2}  & 0 & 0
           \ea\right),
\qquad
\Om^i{_\m}:=\left( \ba{lll}
   \cO_1    &  \cO_3 & \cO_1    \\
   \cO_{-1} &  \cO_1 & \cO_{-1} \\
   \cO_0    &  \cO_2 & \cO_0
           \ea\right),
\ee
\esubeq
where $\om^i$ is the Lie dual of $\om^{ij}$, and $\bar b^i{_\m}$ and
$\bom^i{_\m}$ refer to the background configuration with $\m,B=0$,
representing the massless BTZ black hole. These states are invariant under
the set of restricted local Poincar\'e transformations, defined by the
parameters
\bsubeq\lab{5.2}
\bea
&&\xi^u=\ell U+\cO_2\, ,\qquad
  \xi^r=-r\ell\pd_u U+\cO_0\, ,                                 \nn\\
&&\xi^\vphi=\Phi-\frac{\ell}{r}\pd_\vphi U+\cO_2\, ,            \\
&&\th^+=-\frac{\ell}{r}\pd_\vphi U+\cO_2\, ,\qquad
  \th^-=\frac{r}{2\ell}\pd_\vphi U+\cO_0\, ,                    \nn\\
&&\th^2=\ell\pd_u U+\cO_1\, .
\eea
\esubeq
Here, the functions $U=U(u,\vphi)$ and $\Phi=\Phi(u,\vphi)$ are such that
the combinations $U^\pm=U\pm\Phi$  satisfy the conditions $\pd_\pm
U^\mp=0$, where $x^\pm=u/\ell\pm\vphi$. Since $u=t+\cO_1$ for large $r$,
these conditions define the asymptotic conformal group in 2D.

In spite of certain technical differences between the asymptotic
requirements \eq{5.1} and \eq{B.1}, the corresponding commutator algebras
have the same form. Using the composition law of the restricted Poincar\'e
parameters to leading order, the commutator algebra associated to \eq{5.1}
is found to have the form of two independent Virasoro algebras,
\be
i[\ell^\pm_m,\ell^\pm_n]=(m-n)\ell^\pm_{m+n}\, ,
\ee
where $\ell^\pm_n=\d_0(U^\pm=e^{\pm inx^+})$. The respective central
charges $c^\pm$ will be determined by the canonical methods.

To complete the analysis of the asymptotic conditions, we presented  in
Appendix C an additional set of asymptotic requirements, motivated by the
form of torsion in \eq{4.2}.

\subsection{Canonical generators}\label{sub52}

In order to examine the canonical structure of the quadratic PGT, we use
the \emph{first-order formulation} \cite{x21}, as it leads to a
particularly simple construction of the canonical generator, the form of
which can be found in Eq. (5.7) of Ref. \cite{x7}. In this formulation,
one introduces two new variables, $\t_i$ and $\r_{mn}$, such that their
on-shell values are $\t_i=H_i$ and $\r_{mn}=H_{mn}$. Since the canonical
generator $G$ acts on basic dynamical variables via the Poisson bracket
operation, it is required to be a differentiable phase-space functional.
For a given set of asymptotic conditions, this property is ensured by
adding a suitable surface $\G$ term to $G$, such that $\tG=G+\G$ is
\emph{both differentiable and finite} phase-space functional
\cite{x22,x23}. To examine the differentiability of $G$, we start from the
form of its variation:
\bea
&&\d G=-\int_\S d^2x(\d G_1+\d G_2)\, ,                         \nn\\
&&\d G_1=\ve^{t\a\b}\xi^\mu\left(b^i{_\mu}\pd_\a\d\t_{i\b}
         +\om^i{_\mu}\pd_\a\d\r_{i\b}+\t^i{_\mu}\pd_\a\d b_{i\b}
         +\r^i{_\mu}\pd_\a\d\om_i{_\b}\right)+R\, ,             \nn\\
&&\d G_2=\ve^{t\a\b}\th^i\pd_\a\d\r_{i\b}+R\, .                 \lab{5.4}
\eea
Here, the coherently oriented volume 2-form on the spatial section $\S$ of
spacetime is normalized to $d^2x=dr d\vphi$, the variation is performed in
the set of asymptotic states, $R$ stands for regular terms, and $\r^i$ is
the Lie dual of $\r_{mn}=H_{mn}$, the on-shell value of which reads
\be
H_{ij}=-2a_0\ve_{ijk}b^k-4a_0\ell^2\ve_{ijk}\hat L^k\, ,        \lab{5.5}
\ee
and $\hat L^k$ is the ``symmetrized" Schouten 1-form, $\hat
L_k=L_{(km)}b^m$, see \eq{2.5}.

In what follows, we restrict our considerations by two specific
assumptions that characterize both the OTT black hole and the Vaidya--OTT
solution with torsion:\par
(1) The torsion squared-terms in $L_G$ effectively vanish, that is
$\t_i=0$;\par
(2) ${}^{(5)}R^{ij}=0$.\\
The asymptotic conditions \eq{5.1} imply $\d G_2=R$, so that the surface
term in the improved generator $\tG=G+\G$ is determined by the variational
equations
\bsubeq\lab{5.6}
\bea
&&\d \G=\int_0^{2\pi}d\vphi (\xi^t\d\cE+\xi^\vphi\d\cM)\, ,     \lab{5.6a}\\
&&\d\cE:=\frac{1}{2}\left(\om^{ij}{_t}\d H_{ij\vphi}
                          +\d\om_{ij\vphi}H^{ij}{_t}\right)\, , \lab{5.6b}\\
&&\d\cM:=\frac{1}{2}\left(\om^{ij}{_\vphi}\d H_{ij\vphi}
             +\d\om_{ij\vphi}H^{ij}{_\vphi}\right)\, ,          \lab{5.6c}
\eea
\esubeq
where we used $u=t+\cO_1$, and the boundary $\pd\S$ is parametrized by the
coordinate $\vphi$.

Finding a solution for $\cE$ from the variational equation \eq{5.6b}
demands rather involved considerations, based on the asymptotic conditions
\eq{5.1} and \eq{C.1}. As shown in Appendix C, the surface term for time
translations can be written in the form
\bsubeq\lab{5.7}
\bea
\G[\xi^t]&=&\int_0^{2\pi}d\vphi\,\xi^t\cE\, ,                   \\
\cE&=&\frac{1}{2}\bigl(\om^{ij}{_t} \D H_{ij\vphi}
                    +\D\om^{ij}{_\vphi}\bar H_{ijt}\bigr)
  -\frac{1}{4}\bigl(\D\om^{ij}{_t}\D H_{ij\vphi}
                    -\D\om^{ij}{_\vphi}\D H_{ijt}\bigr)\, ,      \lab{5.7b}
\eea
where $\D X:=X-\bar X$ is the difference between any form $X$ and its
boundary value $\bar X$. On the other hand, equation \eq{5.6c} leads to a
simple surface term for spatial rotations:
\be
\G[\xi^\vphi]=\int_0^{2\pi}d\vphi\,\xi^\vphi\cM\, ,\qquad
  \cM=\frac{1}{2}\om^{ij}{_\vphi} H_{ij\vphi}\, .
\ee
\esubeq
Both $\G[\xi^t]$ and $\G[\xi^\vphi]$ are finite phase-space functionals
(see Appendix C).

The boundary terms for $\xi^t=1$ and $\xi^\vphi=1$,
\be
E=\int_0^{2\pi}d\vphi\,\cE\, ,\qquad M=\int_0^{2\pi}d\vphi\,\cM\, ,
\ee
represent the energy and angular momentum of the system, respectively.
Calculated on the Vaidya--OTT configuration, these expressions take
the values
\be
E=\frac{1}{4G}\left(\m+\frac{1}{4}B^2\ell^2-2K\right)\, ,
\qquad M=0\, .                                                  \lab{5.9}
\ee
The form of $E$ confirms the result \eq{4.7} obtained from the Lagrangian
field equations. In the canonical formalism, the conservation laws for $E$
and $M$ follow from the Poisson bracket algebra of the asymptotic symmetry
\cite{x11}.

The expression for energy defined by equation \eq{5.7b} consists of two
pieces. As shown in Ref. \cite{x15}, the first piece is sufficient to
correctly describe the energy content of a number of solutions in 3D
gravity with/without torsion and topologically massive gravity. However,
when applied to the (Vaidya--)OTT solution, this piece is not sufficient;
in particular, it produces the incorrect coefficient $1/2$ for the $B^2$
term in \eq{5.9}. The second piece in \eq{5.7b} is closely related to the
presence of the $Br$ term in the OTT metric. Thus, our result \eq{5.7b}
represents a generalization of the energy formula used in \cite{x15} to
the (Vaidya--)OTT case.

\subsection{Canonical algebra of asymptotic symmetries}\label{sub53}

The asymptotic symmetry is  described by the Poisson bracket algebra of
the improved generators. Rather then performing a direct calculation, the
form of this algebra can be found by a more instructive method. To show
how it works, we introduce the notation $\tG'=\tG[U',\Phi']$, and
similarly for $\tG''$ and $\tG'''$. Then, according to the main theorem of
Ref. \cite{x23}, one can conclude that the Poisson bracket algebra has the
form
\be
\{\tG'',\tG'\}=\tG'''+C'''\,,                                   \lab{5.10}
\ee
where the parameters of $\tG'''$ are defined by the composition law of the
asymptotic Poincar\'e transformations, and $C'''$ is the central charge
term. In order to calculate $C'''$, one should note that the algebra
\eq{5.10} implies $\d_0'\G''\approx \G'''+C'''$, where $\d_0'\G''$ is
determined by the relations \eq{5.6}, and $C'''$ is identified as the field
independent piece on the right-hand side. Then, going over to the Fourier
modes $L_n^\pm$ of $\tG$, the algebra \eq{5.10} takes the form of two
independent Virasoro algebras,
\be
i\{L^\pm_m,L^\pm_n\}=(m-n)L^\pm_{n+m}+\frac{c^\pm}{12}n^3\d_{m,-n}\, ,
\ee
where the classical central charges are equal to each other, $c^\pm=c$, with
\be
c=\frac{3\ell}{G}\, .
\ee
Thus, the value of $c$ is found to be twice the GR value $c_0=3\ell/2G$.

\section{Concluding remarks}

In this paper, we constructed a Vaidya-like extension of the OTT
black hole as an exact solution of the quadratic PGT in vacuum. The
construction is realized in two steps.

First, we showed that the OTT black hole is a Riemannian vacuum solution
of PGT, provided the coupling constants satisfy certain requirements. The
black hole energy is calculated from the canonical generator for time
translations, the surface term of which is a suitable generalization of
the more standard expression that can be found in Ref. \cite{x14}, see
also Ref. \cite{x15}. The canonical energy $E$ is compatible with the
first law of black hole thermodynamics, in agreement with the equality
of $E$ to the shifted OTT energy \cite{x18}.

Then, following Maeda \cite{x12}, we introduced a Vaidya-like extension of
the OTT black hole; however, this extension is not a Riemannian solution
of PGT in vacuum. To overcome this difficulty, we introduced a suitable
ansatz for the connection possessing a nontrivial torsion content, making
thereby the resulting Vaidya--OTT geometry an exact vacuum solution of
PGT. As far as the asymptotic structure of the Vaidya--OTT solution is
concerned, one should note that: (a) the surface term of the canonical
generator for time translations has the same structure as in the OTT case,
(b) the canonical energy differs from the OTT black hole energy by a
contribution stemming from the torsion, and (c) central charges of the
asymptotic algebra are the same as in the OTT black hole case.

Since the OTT solution is known to exists also for positive or vanishing
$1/\ell^2$ \cite{x10}, most of the present results could be
straightforwardly extended to these sectors.

\section*{Acknowledgements}

This work was supported by the Serbian Science Foundation under Grant No.
171031. The results are checked using the Excalc package of the computer
algebra system Reduce.

\appendix
\section{PGT field equations}
\setcounter{equation}{0}

In this Appendix, we give a brief account of the PGT field equations,
based on Ref. \cite{x7}. The parity-invariant gravitational Lagrangian
$L_G=L_G(b^i,T^j,R^{mn})$ (3-form) is at most quadratic in the torsion
$T^i$ and the curvature $R^{ij}$:
\bea
L_G&=&-\hd(a_0R+2\L_0)
    +T^i\hd (a_1{}^{(1)}T_i+a_2{}^{(2)}T_i+a_3{}^{(3)}T_i)      \nn\\
 && +\frac{1}{2}R^{ij}\left(b_4{}^{(4)}R_{ij}+b_5{}^{(5)}R_{ij}
                       +b_6{}^{(6)}R_{ij}\right)\, ,            \lab{A.1}
\eea
where ${}^{(n)}T^i$ and ${}^{(n)}R^{ij}$ are irreducible components of the
respective field strengths, and $a_0$ is normalized by  $a_0=/16\pi G$. By
varying $L_G$ with respect to $b^i$ and $\om^{ij}$, one obtains the vacuum
field equations that can be written in a compact form as
\bea
\first\quad &&\nab H_i+E_i=0\,,                                 \nn\\
\second\quad &&\nab H_{ij}+E_{ij}=0\, .                         \lab{A.2}
\eea
Here, $H_i:=\pd L_G/\pd T^i$ and $H_{ij}:=\pd L_G/\pd R^{ij}$ are the
covariant momenta:
\bea
&&H_i=2\hd (a_1{}^{(1)}T_i+a_2{}^{(2)}T_i+a_3{}^{(3)}T_i)\,,    \nn\\
&&H_{ij}=-2a_0\ve_{ijm}b^m+H'_{ij}\, ,                          \nn\\
&&H'_{ij}:=2\,\hd\left(b_4{}^{(4)}R_{ij}+b_5{}^{(5)}R_{ij}
                       +b_6{}^{(6)}R_{ij}\right)\, ,            \lab{A.3}
\eea
and $E_i:=\pd L_G/\pd b^i$ and $E_{ij}=\pd L_G/\pd\om^{ij}$ are the
energy-momentum and spin currents:
\bea
&&E_i=h_i\hook L_G-(h_i\hook T^m)H_m
                  -\frac{1}{2}(h_i\hook R^{mn})H_{mn}\, ,       \nn\\
&&E_{ij}=-(b_iH_j-b_jH_i)\, .                                   \lab{A.4}
\eea

In the Riemannian sector ($T^i=0$) with ${}^{(5)}R_{ij}=0$, $H_i$ and
$E_{ij}$ vanish, and the simplified field equations take the form
displayed in \eq{2.7}, with
\bea
&&H_{ij}=-2a_0\ve_{ijm}b^m +\frac{b_6+2b_4}{3}\,R\ve_{ijk}b^k
         -2b_4\ve_{ij}{^m}\ric_{mk}b^k\, ,                      \nn\\
&&E_i=\cL_G\hd b_i-R^{mn}{}_{ik}b^kH_{mn}\, .                   \lab{A.5}
\eea
Here, we used $L_G=\cL_G\,\heps$, and $\heps$ is the volume 3-form.

\section{Asymptotic conditions for the OTT black hole}
\setcounter{equation}{0}

The action of the AdS Killing vectors on the OTT black hole configuration,
described in section 2, leads to the associated asymptotic conditions that
are \emph{relaxed} with respect to the Brown-Henneaux ones:
\bsubeq\lab{B.1}
\be
b^i{_\m}=\bar b^i{_\m}+B^i{_\m}\,,\qquad
B^i{_\m}:=\left( \ba{lll}
         \cO_0  & \cO_3  & \cO_0  \\
         \cO_1  & \cO_2  & \cO_1  \\
         \cO_0  & \cO_3  & \cO_0
               \ea\right)\,,
\ee
and
\be
\om^i{_\m}=\bar\om^i{_\m}+\Om^i{_\m}\,,\qquad
\Om^i{_\m}:=\left( \ba{lll}
   \cO_0  &  \cO_3 & \cO_0   \\
   \cO_1  &  \cO_2 & \cO_1   \\
   \cO_0  &  \cO_3 & \cO_0
           \ea\right)\, .
\ee
\esubeq
Here, $\om^i$ is the Lie dual of $\om^{ij}$, and $\bar b^i{_\m}$ and
$\bom^i{_\m}$ refer to the AdS background (with $B=\mu=0$). These
conditions are invariant under the asymptotic Poincar\'e transformations,
defined by the set of restricted local parameters $(\xi^\m,\th^i)$ that
can be found in Ref. \cite{x20}. The conditions \eq{B.1} are a PGT
generalization of those discussed in \cite{x10}.

Following the procedure described in section 5, one can find the conserved
charges of the OTT black hole, the energy $E$ and the angular momentum
$M$. Moreover, the canonical algebra of the asymptotic symmetry is
represented by two independent Virasoro algebras with equal central
charges $c^\mp=c$. The values of $E,M$ and $c$ are given in subsection
\ref{sub23}.

\section{Refined asymptotic conditions}
\setcounter{equation}{0}

Equation \eq{4.2} implies that the Vaidya--OTT solution has only one
nonvanishing component of torsion: $T^2{}_{u\vphi}=K$. Clearly, this
property is not valid on the whole set of asymptotic states. In order to
ensure finiteness of the improved canonical generators, we find it
necessary to make further restrictions of the asymptotic conditions
\eq{5.1} by demanding the highest order terms in $T^i{}_{\m\n}$ to vanish:
\bsubeq\lab{C.1}
\bea
&&T^+{}_{u\vphi}:\quad
   r\left(\Om^+{_u}+\frac{1}{\ell^2}B^+{_\vphi}\right)
  +(\Om^2{_\vphi}+B^2{_u})=\cO_1\, ,                            \nn\\
&&T^+{}_{ur}:\quad \frac{r}{\ell^2}B^+{_r}+\Om^2{_r}
                   +\frac{1}{r}B^+{_u}=\cO_3\, ,                \nn\\
&&T^+{}_{r\vphi}:\quad
  r\Om^+{_r}+B^2{_r}+\frac{1}{r}B^+{_\vphi}=\cO_3\, ,           \\
&&T^-{}_{u\vphi}:\quad
    r\left(\Om^-{_u}+\frac{1}{\ell^2}B^-{_\vphi}\right)
    +\frac{r^2}{2\ell^2}(\Om^2{_\vphi}+B^2{_u})=\cO_0\, ,       \nn\\
&&T^-{}_{ur}:\quad \frac{r}{\ell^2}B^-{_r}
   +\frac{r^2}{2\ell^2}\Om^2{_r}-\Om^2{_u}
   -\frac{1}{r}B^-{_u}=\cO_1\, ,                                \nn\\
&&T^-{}_{r\vphi}:\quad
  \Om^2{_\vphi}+\frac{r^2}{2\ell^2}B^2{_r}
  +r\Om^-{_r}+\frac{1}{r}B^-{_\vphi}=\cO_1\, ,                  \\
&&T^2{}_{u\vphi}:\quad
   \frac{r^2}{2\ell^2}(\Om^+{_\vphi}+B^+{_u})
  -(B^-{_u}+\Om^-{_\vphi})=\cO_0\, ,                            \nn\\
&&T^2{}_{ur}:\quad
  \frac{r^2}{2\ell^2}\Om^+{_r}-(\Om^-{_r}+\Om^+{_u})=\cO_2\, ,  \nn\\
&&T^2{}_{r\vphi}:\quad
   \frac{r^2}{2\ell^2}B^+{_r}+\Om^+{_\vphi}-B^-{_r}=\cO_2\, .
\eea
\esubeq

Now, we use the asymptotic conditions \eq{5.1} and \eq{C.1} to derive the
surface terms \eq{5.7} and prove their finiteness. First, we show that
$\cE$ satisfies the variational equation \eq{5.6b}:
\bea
\d\cE&=&
\frac{1}{2}\left(\om^{ij}{_t}\d H_{ij\vphi}
                 +\d\om_{ij\vphi}H^{ij}{_t}\right)              \nn\\
  && +\frac{1}{4}\left(\d\om^{ij}{}_t\D H_{ij\vphi}
   -\D\om^{ij}{}_t\d H_{ij\vphi}-\d\om^{ij}{}_\vphi\D H_{ijt}
   +\D\om^{ij}{}_\vphi\D H_{ijt}\right)                         \nn\\
&=&\frac{1}{2}\left(\om^{ij}{_t}\d H_{ij\vphi}
                     +\d\om_{ij\vphi}H^{ij}{_t}\right)+\cO_1\, ,
\eea
Next, we prove that the surface term for time translations is finite:
\bea
\cE&=& 2a_0\left(\frac{r^2}{2\ell^2}\Om^+{_\vphi}
       +\Om^-{_\vphi}-r\Om^2{_u}\right)+\cO_0                   \nn\\
&=& -a_0\frac{r^3}{\ell^2}\left(\frac 1r B^+{_u}+\frac r{\ell^2}B^+{_r}
       +\Om^2{_r}\right)+\cO_0=\cO_0\, .
\eea
Finally, we derive the finiteness of the surface term for spatial rotations:
\bea
\cM&=&-a_0\ve^{imn}\om_{mn\vphi}\Bigl(b_{i\vphi}
     +2\ell^2L_{(ij)}b^j{_\vphi}\Bigr)                          \nn\\
&=&2a_0\left(rB^2{_u}-B^-{_\vphi}
  -\frac{1}{2}\frac{r^2}{\ell^2}B^+{_\vphi}\right)
  -4a_0\ell^2\left(r\ric_{(+2)}+\frac{r^2}{2\ell^2}\ric_{(-2)}\right)
  +\cO_0                                                        \nn\\
&=&\cO_0\, .
\eea

\end{document}